
\magnification=1200
\hsize 16 true cm
\baselineskip=15pt

\centerline{\bf THE VELOCITY OF THE EMITTING PLASMA OF}
\centerline{\bf THE SUPERLUMINAL GALACTIC SOURCE GRS 1915+105}
\vskip 1 true cm
\centerline{ Gianluigi Bodo and Gabriele Ghisellini}
\vskip 0.5 true cm
\centerline{Osservatorio Astronomico di Torino}
\centerline{Strada Osservatorio, 20 10025 Pino Torinese, Italy}
\vskip 1 true cm
\centerline{\bf ABSTRACT}
\vskip 0.5 true cm
\noindent
We calculate the flux and size ratio of radio emitting features moving in
antiparallel directions at relativistic speeds, taking into account that
the pattern we observe may move at a velocity different from the one of
the emitting plasma.
Both velocities can be determined in sources in which both the
approaching and the receding features are observed.
This is the case of the galactic superluminal source GRS
1915+105, for which the pattern speed (responsible for the
apparent superluminal velocity) was found to be 0.92$c$.
For this source we find that the velocity of the plasma is 0.73$c$.
The found velocity helps decreasing the estimate of the associated
kinetic power.
Since intrinsically identical knots will be observed
to have different sizes, high resolution radio observations
of GRS 1915+105 will be a test for the proposed difference
of the pattern and plasma velocities.

\noindent
{\it Subject headings:} Relativity -- Radio continuum: stars ---
Stars: GRS 1915+105

\vskip 1 true cm
\centerline{\bf 1. INTRODUCTION}
\vskip 0.5 true cm

\noindent
Two oppositely directed blobs of radio emission have been recently
observed to move at relativistic speeds in GRS 1915+105
(Mirabel \& Rodriguez, 1994, hereinafter MR94).
The apparent velocities are $\beta_a=1.25\pm$ 0.15 and 0.65$\pm$0.08:
one of the blobs is therefore superluminal.
This is the first example that the violent phenomena occuring in
active galactic nuclei (AGNs) may be reproduced on a much smaller
scale also by objects in our own galaxy.
Due to the proximity of this source, probably located at 12.5 Kpc
from us, we have a unique oppurtunity to study relativistic effects
in much greater depth than what is possible for AGNs.
Indeed, GRS 1915+105 is the first example of superluminal source
in which we can observe a $pair$ of blobs going away from a stationary
center.
This has already allowed MR94 to determine $both$ the real velocity
of the moving patterns $and$ the angle of motion with respect to the line
of sight.
Using the derived `true' speed ($\beta=0.92\pm 0.08$) and viewing angle
($\theta=70^\circ\pm 2^\circ$), MR94 were also able to estimate the
total emitting mass, the kinetic energy of its bulk motion, and
the corresponding kinetic power, which turned out to be worryingly
large: $L_{kin}\sim 10^{41}$ erg s$^{-1}$.
Furthermore, the knowledge of $\beta$ and $\theta$ immediately
lead to the prediction of the expected frequency shift of any emission line,
whose detection would be, of course, very important to confirm our
ideas about beaming models.

In this $Letter$ we will argue that the `true' velocity of the emitting
plasma can (and very likely is) lower than the velocity of the moving
pattern responsible for the apparent superluminal speed.
This will partly ease the problem in explaining the large
kinetic power and allows to make a more detailed prediction
of the frequency shift of the emission line by the outflowing matter.
The above results are largely model independent.
We also calculate the apparent sizes of the approaching and
receding knots,
and discuss how their ratio can give further independent
informations on the pattern and plasma velocities.

\vskip 1 true cm
\centerline{\bf 2. OPPOSITELY MOVING BLOBS}
\vskip 0.5 true cm
{\it 2.1 FLUX RATIO}
\vskip 0.5 true cm

\noindent
In this section we offer a simple derivation of the expected
flux density from two blobs moving in antiparallel directions.
The pattern of each blob is moving at some velocity $\beta_s$,
while the velocity of the fluid inside each blob is moving
with a different velocity $\beta_F$.
All velocities are measured in the observer frame, and the angle
$\theta$ is between the line of sight and the velocity of
the approaching blob.
As a special case, we may have $\beta_s=\beta_F$ (the plasmon case),
for which the observed optically thin flux ratio is
(for power law emission $F(\nu)\propto \nu^{-\alpha}$)

$$
{ F_a(\nu) \over F_r(\nu)} \, =\,
\left( { 1+\beta\cos\theta \over 1-\beta\cos\theta} \right)^{3+\alpha}
, \qquad
(\beta=\beta_s=\beta_F)
\eqno(1)
$$
here the subscripts $a$ and $r$ stand for appraching and receding,
respectively.

In the more general case of $\beta_s\ne \beta_F$, we have to integrate
the emissivity (as it is in our observer frame) over the observed volume.
It is of course better to express the emissivity in a frame comoving
with the emitting plasma (comoving quantities will be denoted
by a prime), using

$$
j_{a,r}(\nu) \, = \,
\left[ { 1\over \gamma_F(1\pm \beta_F\cos\theta)} \right]^{2+\alpha}
j\prime_{a,r}(\nu) \, \equiv \,
\delta_{f\, a,r}^{2+\alpha}j\prime_{a,r}(\nu)
\eqno(2)
$$
where $\delta_F$ is the Doppler factor of the fluid.
A simple way to derive the above transformation is to write
$j(\nu)$ as the number of photons of energy $h\nu$ emitted
per unit time, frequency, solid angle and volume, i.e.
as $\propto dNh\nu/(dt d\nu d\Omega dV)$ and Lorentz
transforming all the quantities (i.e. Rybicki \& Lightman, 1979).

What is the observed volume?
When we make a flux measurement (or a map) we collect photons
arriving at the same time to the detector.
For extended objects, therefore, these photons leaved the source
at different times (Terrel, 1960).
For the approaching blob, the photons emitted at the back of the
blob have to leave at earlier times than the photons emitted at the
front.
This time interval corresponds also to different locations of the
blob, whose front moves with the pattern velocity $\beta_sc$.
The reverse is true for the receding blob.
The ratio of the approaching and receding volumes is
$$
{ V_a \over V_r }\, = \,
{ 1+\beta_s\cos\theta \over 1-\beta_s\cos\theta}
\eqno(3)
$$
Therefore the ratio of the observed fluxes is

$$
{ F_a(\nu) \over F_r(\nu)} \, =\,
{ 1+\beta_s\cos\theta \over 1-\beta_s\cos\theta}
\left( { 1+\beta_F\cos\theta \over 1-\beta_F\cos\theta} \right)^{2+\alpha}
\eqno(4)
$$
Special cases are:

i) $\beta_s =\beta_F$. Then equation (4) reduces to equation (1).

ii) $\beta_s=0$. This is the case of a $stationary$ jet,
whose volume is fixed in the observer frame.

iii) $\beta_F=0$. This is the case of antiparallel perturbations
travelling along the jet, powering different portions of
plasma at rest (or slowly moving).
Even in this case the flux ratio is greater than unity,
and can reach large values for small angles.

The same result (i.e. equation 4) can be derived by using the formalism
of Lind \& Blandford (1985), in particular using their equation (2.12b),
but we think that our derivation is simpler and readily applicable to
the case of superluminal sources in which both approaching and receding
blobs are observed, as is the case of GRS 1915+105.

\vskip 1 true cm
{\it 2.2 SIZE RATIO}
\vskip 0.5 true cm

\noindent
Two opposetely moving blobs, identical in their rest frame,
will have different sizes in the observer frame.
For illustration, consider the case of two cilinders of heigth $l$
and section diameter $d$, measured in its proper frame, moving
at velocities $\pm \beta_s$ along a direction which makes an angle
$\theta$ with the line of sight. When we compute its projected length
seen by the observer, we have to take into account aberration,
time delay and Lorentz contraction effects. We note that, as in
the previous discussion,  the projected length is that defined by the
photons collected at the same time by the observer, which, therefore,
leaved the source at different times.  The total length observed is made
by two contributions, the projected heigth and the projected section
of the cylinder.  The first one, being parallel to the velocity is
modified by Lorentz contraction and time delay and becomes
$$
l^{obs} = l \sin \theta \; {1 \over \gamma_s} \; { 1 \over {1 - \beta_s \cos
\theta}}  = l \delta_s \sin \theta
\eqno(5)
$$
For the second one we have only to consider time delay effects, and we
have
$$
d^{obs} \, = \,  { \vert \cos\theta -\beta_s \vert \over 1-\beta_s\cos\theta }
\, \, d
\eqno(6)
$$
We note that, when $\beta_s < \cos \theta$, we see the face of the
cylinder closer to us, when   $\beta_s = \cos \theta$, the projected size
of the section of the cylinder is zero and we see only its lateral
surface, while, when  $\beta_s > \cos \theta$, we see its farther face.

The total projected length is $l^{obs} + d^{obs}$ and the ratio between
that observed for the approaching cylinder and that observed for the
receding one is
$$
R =  { {\sin \theta + x \gamma_s \vert \beta_s - \cos\theta \vert} \over
{\sin \theta + x \gamma_s  (\beta_s + \cos\theta)} }
{ {1 + \beta_s \cos\theta} \over { 1 - \beta_s \cos\theta} }
\quad ,
\eqno(7)
$$
where $x = d/l$.
The same result could be obtained considering the invariance
of the perpendicular separation between two parallel light rays
(see e.g. the discussion after equation 2.6 of Lind \& Blandford 1985).
Using this method the observed total length can be obtained by the
following procedure:
i) calculate the aberration angle relative to the
observer and the proper frame;
ii) in the proper frame, consider the direction of the
`de--aberrated' rays;
iii) project the cylinder (in the proper frame) in a plane
perpendicular to the this direction.
This is the projected shape seen by the observer.
Using this method is immediately apparent that the maximum length of a
fast moving object cannot be larger than the maximum length seen in the
proper frame projecting the object in different directions.
(i.e. the maximum lenght of a cylinder is $\sqrt{d^2+l^2}$).

In a real situation we can imagine that the cylindrical regions observed
are associated to  shocks and correspond to the regions of emitting
post--shock material. Their length will therefore be determined by the
cooling time of the post--shock material and will be in general a
function of the frequency of observation.
Defining the post--shock plasma velocity, in the shock frame, as
$\beta_{F}^s$ and its cooling time, in the plasma proper frame, as
$t_{cool}^F(\nu^F)$, we can compute the cooling length in the shock frame as
$$
l^s_{cool} = c t^F_{cool}(\nu^F) \gamma^s_{F} \beta^s_{F}
$$
In the previous and following expressions the superscripts $F$, $s$
denote quantities measured respectively in the shock and plasma frames.
To find how the cooling time depends on frequency, we must assume
a given emission mechanism.
For synchrotron $t_{cool} (\nu) \propto \nu^{-1/2}$ which yields
$t^F_{cool} (\nu^F) = t^F_{cool} (\nu) \delta_F^{1/2}$, where
$\nu$ is the observed frequency.
Expressing now $\beta^s_{F}$ and $\gamma^s_{F}$ in term of the plasma
and shock velocities seen by the observer and using the observed
frequency, we obtain the following expression for the quantity $x$
appearing in Eq. (7):
$$
x (\beta_s, \beta_F, \theta, \nu) =
{ d \over {c t^F_{cool} (\nu) \delta_F^{1/2} \gamma_s \gamma_F (\beta_s
- \beta_F) }  }
\eqno(9)
$$

Special cases are:

i) $\beta_s=\beta_F$: in this case equation (7) yields
$R= \sqrt{(1+\beta_s)/(1-\beta_s)}\vert \cos\theta-\beta_s \vert /
(\cos\theta+\beta_s)$, which is always less than unity.

ii) $d \ll c t_{cool}^F$ : in this limiting case $R=(1+\beta_s\cos\theta)/
(1-\beta_s\cos\theta)$,
indipendently on the fluid velocity, reaching very large values
for small viewing angles.

iii) $\beta_F=0$: even in this case we expect $R$ to be, in general,
different from unity, especially at small viewing angles.

\vskip 1 true cm
\centerline{\bf 3. APPLICATION TO GRS 1915+105}
\vskip 0.5 true cm

\noindent
As measured by MR94, the flux ratio of approaching and receding
blobs in GRS 1915+104, calculated at equal separations from the
center (hence when the blobs have the same age) is $\sim 8\pm 1$,
inconsistent with the value ($F_a/F_r\sim 12$ ) expected in the case
of equal pattern and fluid velocities.
Using our equation (4) with $\beta_s=0.92$ and $\theta=70^\circ$,
we derive $\beta_F=0.73$ assuming a flux ratio
equal to 8 (it becomes 0.664 and 0.787 with flux ratios equal to
7 and 9, respectively), corresponding to a Lorentz factor
$\gamma_F=1.46$.

Any line emitted by the moving fluid would be shifted in frequency
by $\delta_{F,a}=0.911$ (approaching fluid) and
by $\delta_{F,r}=0.547$ (receding fluid), which are significantly
different by the values found using $\beta=0.92$ and quoted in MR94.

With the derived value of $\beta_F$ we can predict the size ratio $R$ as
a function of $x(\nu)$, which is unfortunately unknown.
However, for $x(\nu)=$0, 1 and 100 we have $R=$ 0.92, 0.84 and 0.68,
respectively.
Even if the precise value of $x(\nu)$ is not known, a well defined
behaviour of $R$ is predicted: $R$ should decrease by observing at
different, increasing, frequencies, which correspond to shorter
cooling times and therefore at larger $x(\nu)$.

Also the estimate of the kinetic power will be significantly affected.
In fact, in order to estimate the total mass producing the synchrotron
emission we observe, we have to correct the observed luminosities
and frequencies with the correct Doppler factor.
Following MR94, we may approximate the energy distribution of the
emitting particles with a monoenergetic function at $\gamma=10^3\gamma_3$,
a tipical value to produce an observed frequency of
$10 \nu_{10}\delta$ GHz.
This fixes a fiducial value for the magnetic field
$B=2.7\times 10^{-3} \nu_{10}\delta/\gamma_3^2$.
Charge neutrality may be provided by protons or positrons.
Since electron and positrons would have the same large $\gamma$
while protons are very likely not relativistic, the hypothesis of a pure
pair plasma would only slightly affect (factor 2) the mass estimate.

With the values of $\delta$ derived above, we obtain a new
estimate for the total mass of $M=7.6 \times 10^{24}$ g, a factor 3
lower than MR94.
The kinetic luminosity is $L_{kin}=2.6\times 10^{40}$ erg s$^{-1}$,
a factor $\sim 4$ lower than MR4.

\vskip 1 true cm
\centerline{\bf DISCUSSION}
\vskip 0.5 true cm

\noindent
Superluminal sources in which both the approcahing and the receding blobs
are observed offer the oppurtunity to estimate both the velocity of the
pattern and the velocity of the emitting plasma.
For the recently discovered galactic source GRS 1915+105 the estimated
plasma velocity is $\beta_s=0.73\pm 0.06$,
significantly lower than the pattern speed $\beta_s=0.92\pm 0.08$.

This velocity difference is suggestive of a shock propagating along the jet
with velocity $\beta_s$ and postshock velocity $\beta_F$.
The velocity difference found would indicate a compression ratio lower
than that expected for strong relativistic shocks in a non magnetized
medium. Our result is therefore indicative that the magnetic field is
likely to be important for the dynamics of the flow observed in this
source, however more precise conclusions are precluded by the
uncertainties in the relatistic shock theory  (Appl \& Camenzind 1988;

If the moving knots are not spheres, as our analysis implies,
than their apparent, projected shapes are modified by
Lorentz contraction and by the different travelling times of
the photons emitted in different part of the source.
This modifications are different for the approaching and the receding
knots, in such a way that two intrinsically (when at rest) equal
knots are observed to have a different total projected size.

With sufficient angular resolution, the approaching and receding blobs
of GRS 1915+105 should not be observed as circles of equal diameters,
but with elongated patterns of different maximum extension.
The ratio of their sizes can confirm or not the plasma velocity derived
from the flux ratio.

As mentioned, any line emitted by the plasma should be observed
shifted in frequency, immediately giving (for a given angle) the
corresponding velocity.
Formation of lines could however be strongly inhibited
in a too hot gas.
An attractive alternative is the annihilation line, expected
if part of the jet material is made by electron--positron pairs.
Note that the source 1E1740--294, close to the galactic center, is
thought to have jets carrying electron--positron pairs, to explain the
variable annihilation line observed by the SIGMA satellite (Bouchet et
al. 1991, Sunyaev et al. 1991, Ramaty et al. 1992).
GRS 1915+105 may be similar.
We may very roughly calculate an upper limit to the annihilation line
flux assuming that all the particles emitting in the radio, once cooled,
annihilate before escaping the source.
These particles, of initial energy $\gamma_0$, emit a fraction
$(\gamma_0-1)/\gamma_0$ of their total energy as radio emission,
and a fraction $1/\gamma_0$ as annihilation radiation.
Therefore $F_{ann}\sim F_{radio}/\gamma_0$ corresponding to a photon
flux $\dot N\sim 10^{-7} \nu_{10}F_{Jy}(\nu_{10})/\gamma_0$
[photons cm$^{-2}$ s$^{-1}$].
This limit can be increased if only a fraction of the existing pairs have
been accelerated to high energies to radiate in the radio.
In this case one can obtain another upper limit by considering that
the nucleous of the source transforms a fraction $\xi$ of its
hard X--ray luminosity into pair rest mass.
The maximum value of $\xi$ is thought to be around 10 per cent
(Svensson, 1987) for non thermal plasmas, and much less
for thermal ones.
In this case, the estimated limit for the annihilation photon flux is
$\dot N \sim 6.5 \times 10^{-3} \xi L_{38}$
[photons cm$^{-2}$ s$^{-1}$], where $L_{38}$ is the luminosity in
hard X--rays in units of $10^{38}$ erg s$^{-1}$.

\vskip 1 true cm
\centerline{\bf REFERENCES}
\vskip 0.5 true cm

\parindent=0 pt
\everypar={\hangindent=2.6pc}

Bouchet L. et al., 1991, ApJ, 383, L45

Lind, K.R. \& Blandford, R.D., 1985, ApJ, 295, 358

Mirabel, I.F. \& Rodriguez, L.F., 1994, Nature, 371, 46 (MR94)

Ramaty, R., Leventhal, M. Chan, K.W. \& Lingenfelter, R.E., 1992, ApJ,
392, L63

Rybicki, G.B. \& Lightman A. P., 1979, {\it Radiative Processes
in Astrophysics}, ed John Wiley \& Sons (New York)

Sunyaev, R. et al., 1991, ApJ, 383, L49

Svensson, R., 1987, MNRAS, 227, 403

Terrel, J., 1959 , Phys. Rev., 116, 1041

\bye